%% file: sample-sigconf.tex
\documentclass[sigconf]{acmart}

\usepackage{booktabs} 

\setcopyright{rightsretained}

\acmConference[SIGSPATIAL'17]{ACM SIGSPATIAL conference}{November 2017}{Redondo Beach, California USA} 
\acmYear{2017}
\copyrightyear{2017}

\begin{document}
\title{Assessing verticalization effects on urban safety perception}


\author{Ricardo Barros Louren\c{c}o}
\orcid{0000-0002-4158-3244}
\affiliation{%
  \institution{University of Chicago}
  \streetaddress{1100 E. 58th Street - Ryerson Physical Laboratory}
  \city{Chicago} 
  \state{Illinois} 
  \postcode{60637}
}
\email{rlourenco@uchicago.edu}

\begin{abstract}
We describe an experiment with the modeling of urban verticalization
effects on perceived safety scores as obtained with computer vision on \textit{Google Streetview}
data for New York City. Preliminary results suggests that for smaller buildings (between one and seven floors),
perceived safety increases with building height, but that for high-rise buildings, 
perceived safety decreases with increased height. 
We also determined that while height
contributing for this relation, other zonal aspects also influences the perceived safety scores, 
suggesting spatial structuring also influences such scores.
\end{abstract}



\maketitle

\input{samplebody-conf}

\bibliographystyle{ACM-Reference-Format}
\bibliography{sample-bibliography} 

\end{document}

%% file: samplebody-conf.tex
\section{Introduction}
Urban designers argue that cities should have a ``human scale.'' 
For example,  Teng~\cite{Teng2012} argues 
that buildings should be at most five stories high, as this is ``the maximum height for people to build the visual and emotional relations with the urban space which is attached to the building.'' 

Inspired by these arguments,
we decided to investigate whether one can quantify the impact of building height
on human experiences of urban spaces.
To this end, we combine building data from a City of New York data portal with a
dataset, Streetscore, which associates perceived safety index values
(q-scores) with \textit{Google Streetview} images.
Our goal was to see how perceived safety scores may relate to building heights, 
and if this relationship changes in geographic space.

%

\section{Methodology}
We used GeoDa~\cite{anselin2006geoda} exploratory spatial data analysis (ESDA) capabilities, such as the ability to link and brush values present simultaneously in regression graphs and in their related maps,
to evaluate the relation between building heights and Streetscore q-scores. 
We also made use of GeoDa routines to perform Local Analysis of Spatial Association (LISA)~\cite{anselin2014modern} to evaluate spatial structuring.

\subsection{Dataset Descriptions}

The Streetscore dataset for NYC comprises 322,386 \textit{Google Streetview} images,
each with an associated safety score obtained via crowdsourcing~\cite{6910072}.
Individuals were repeatedly shown pairs of images and asked to indicate which depicted a safer area.
Multiple such evaluations were then combined to generate a georeferenced global rank (q-score), 
such that lower values represent the perception of less safe areas and higher scores a more secure area. 

We obtained, from the City of New York data portal, 
digital footprints and building heights for 1,082,349 buildings in New York City as of 2014.
We also retrieved polygons for 195 NYC neighborhood areas, for use when aggregating q-scores.

\subsection{Data Transformation}
We ingested the Streetscore and building
datasets into a PostgreSQL environment, with PostGIS module loaded. 
We then used QGIS and GeoDa~\cite{anselin2006geoda} to perform
a series of data transformations in order to integrate the 
Streetscore q-index points and building footprint polygons, as follows. 

First, we used the Scikit-learn~\cite{pedregosa2011scikit} k-nearest neighbor (KNN) library to determine the
30 nearest q-score neighbors for each building polygon centroid,
which we then interpolated to obtain an estimated q-score for that building.
We chose to work with 30 neighbors/samples to avoid an eventual sample bias, if the population is skewed.
We interpolate based on nearest neighbors rather than a fixed distance such as Euclidean or Manhattan distance both to obtain our desired 30 samples and to
avoid too much overlapping on space, possibly increasing multicolinearity. 


Next, we used the QGIS spatial aggregation feature to aggregate the per-building
interpolated q-scores to obtain average q-score values for each of the 195 New York City neighborhood areas.
In addition, because the GeoDa LISA function requires spatially continuous data, 
and the community areas are not spatially continuous, 
we used a Voronoi tessellation method from GeoDa to process the
community area centroids 
to generate a contiguous space.

Finally, we used the GeoDa LISA function to obtain a cluster map. 
For this we used a queen weight matrix with spatial contiguity of order one, 
to process the obtained Voronoi multi-polygon layer.

\section{Results and Conclusion}
\begin{figure}
\includegraphics[height=2.125in, width=3.0in]{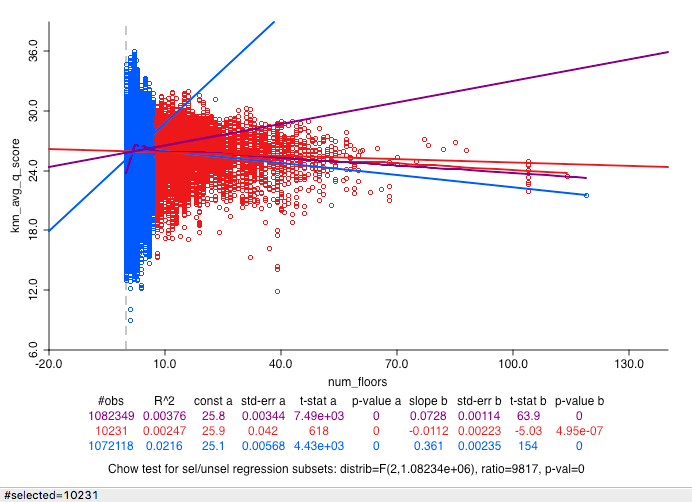}
\caption{Lower buildings (blue), higher buildings(red) and their linear regression and LOWESS regression lines}
\label{fig:lowslope}
\end{figure}

\begin{figure}
\includegraphics[height=1.597in, width=2.111in]{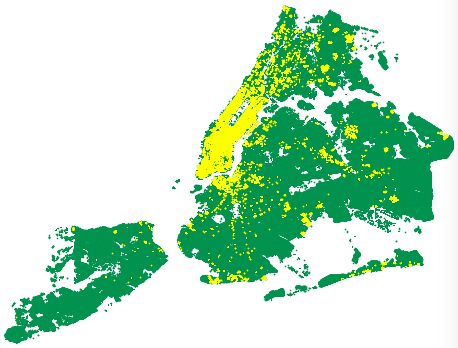}
\caption{This linked map of Figure \ref{fig:lowslope} relates the printed yellow zone with the red (high-rise) areas of the linked graph. There is correlation of the distribution of high-rise areas in traditionally privileged zones, but also in other areas of the city, showing spatial structuring}
\label{fig:lowslope_map}
\end{figure}

Figure \ref{fig:lowslope} shows a least-squares regression of \textit{number of floors} as independent variable and  \textit{averaged q-score} as dependent variable. 
The coefficient of determination (R$^2$) is low: less than five percent. 
This result does not necessarily imply that such factor is irrelevant on the analysis, but its contribution is small, and probably adds to other factors contribution to the dependent variable mapping. 
Another aspect is that such mapping could be non-linear in the geographic space, and this method would not be suitable for proper evaluation.
Even considering such constraints, it is possible to see that there are two different groups of mappings, one in a lower tier of number of floors (smaller than eight), which accounts for around 90\% of the buildings in NYC, revealing differential structuring as revealed on the linked map for the same dataset (Figure \ref{fig:lowslope_map}), where higher buildings are preferably in Manhattan, and in some spots across the city.

\begin{figure}
\includegraphics[height=1.681in, width=3.0in]{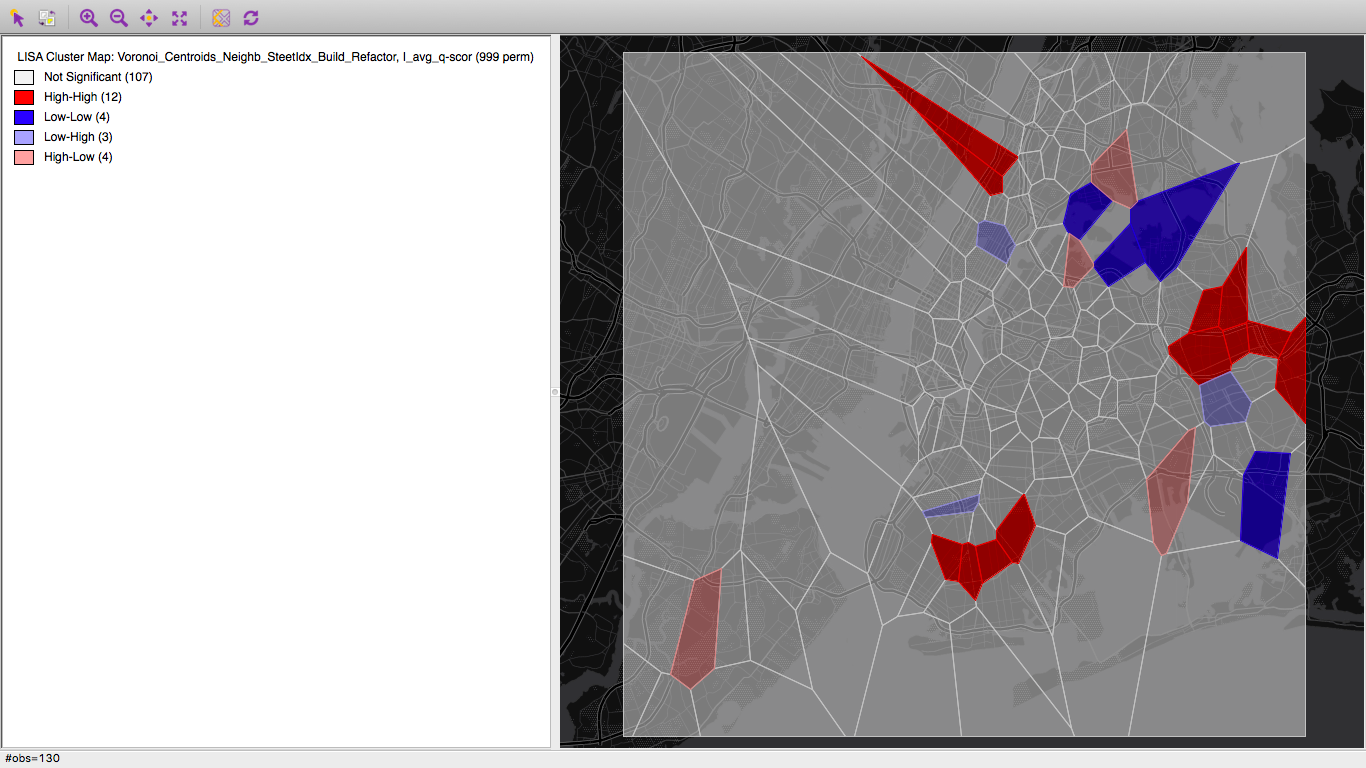}
\caption{LISA Cluster map showing High-High areas in dark red, Low-Low areas in dark blue, Low-High areas in light blue, and High-Low areas in light red. Blue and red areas have $>$95\% significance}
\label{fig:LISA_Cluster}
\end{figure}

The LISA Cluster Map (Figure \ref{fig:LISA_Cluster}) allows us to assess spatial issues.
The colored clusters account for the areas that have more than 95\% statistical significance. 
The red areas (high-high) are zones that have high q-index into a region that the adjacent polygons have also high q-indexes, 
such as Hamilton Heights and Manhattanville in Manhattan, and Fresh Meadows, Auburndale, Flushing, Flushing Heights, and Bellerose in Queens. 
The blue (low-low) sectors are the ones that showed low score clusters, in a region of low scores (even if the surroundings have lower statistical significance), such as regions in the Bronx and Queens. 
A low-high region was detected in Queens (Jamaica), which exactly points the difference on perceived safety, suggesting that this region is less safe than its surroundings.

With this work it was possible to see some spatial structuring regarding q-scores, also regarding the building height effects, suggesting that on a lower building scale, the safety perception is positively correlated until eight floors. 
When equal or higher then eight floors, it suggests a negative correlation, results which are aligned with the assumption made by architects. 
Considering that several other factors account for the welfare perception of citizens~\cite{Teng2012}, it would be beneficial to add other data sources, such as US Census, and other administrative data derived from commerce, to investigate whether they also contribute to a general safety perception.

\begin{acks}
The work is supported in part by NSF Decision Making Under Uncertainty program award 0951576. 
The author thanks Professors Luc Anselin and Ian Foster (his PhD advisor) for feedback.

\end{acks}